\documentclass[english,prl, twocolumn,notitlepage, superscriptaddress]{revtex4-1}
\usepackage{amsmath}
\usepackage{graphicx}

\makeatletter
\usepackage[usenames,dvipsnames]{xcolor}
\usepackage[colorlinks=true,citecolor = MidnightBlue,
urlcolor = ForestGreen, linkcolor=BrickRed]{hyperref}

\makeatother

\usepackage{babel}
\begin{document}

\title{How isolated is enough for an ``isolated'' system in statistical
mechanics?}

\author{Hui Dong}
\email{huidong@tamu.edu}

\affiliation{Institute of Quantum Science and Engineering, Texas A\&M University,
College Station, Texas}

\author{Da-wei Wang}

\affiliation{Institute of Quantum Science and Engineering, Texas A\&M University,
College Station, Texas}

\author{M. B. Kim}

\affiliation{Institute of Quantum Science and Engineering, Texas A\&M University,
College Station, Texas}
\begin{abstract}
Irreversible processes are frequently adopted to account for the entropy
increase in classical thermodynamics. However, the corresponding physical
origins are not always clear, e.g. in a free expansion process, a
typical model in textbooks. In this letter, we study the entropy change
during free expansion for a particle with the thermal de Broglie wavelength
($\lambda_{T}$) in a one-dimensional square trap with size $L$.
By solely including quantum dephasing as an irreversible process,
we recover classical result of entropy increase in the classical region
($L\gg\lambda_{T}$), while predict prominent discrepancies in the
quantum region ($L\ll\lambda_{T}$) because of non-equilibrium feature
of trapped atoms after expansion. It is interesting to notice that
the dephasing, though absent in classical system, is critical to clarify
mysteries in classical thermodynamics.
\end{abstract}
\maketitle
\narrowtext

Quantum coherence, crucial to current investigations of quantum information,
quantum computation\cite{Nielsen2004unknown_,Bennett2000Nature404_247,Ladd2010Nature464_45},
and quantum simulations \cite{Bloch2008Nature453_1016,Georgescu2014Rev.Mod.Phys.86_153},
is also capable of producing surprising effects in thermodynamics
and statistical mechanics \cite{Asher1995book,Scully2003Science299_862}.
For example, a heat engine, powered by an ensemble of trapped atoms
with quantum coherence, could surpass the Carnot efficiency \cite{Scully2003Science299_862,Scully2002643_83},
yet with no violation of the Second Law of Thermodynamics \cite{Scully2002643_83,Quan2006Phys.Rev.E73_36122}.
Importantly, recent progresses in cold atom physics have allowed real
experimental demonstrations of many quantum thermodynamic phenomena
\cite{Meyrath2005Phys.Rev.A71_41604,Catani2009Phys.Rev.Lett.103_140401,Kim2011Phys.Rev.Lett.106_70401,Gaunt2013Phys.Rev.Lett.110_200406,Rossnagel325},
at a region where the size of the trap is comparable with the thermal
de Broglie wavelength \cite{pethick2002bose}. Within these trapped
atomic systems \cite{Meyrath2005Phys.Rev.A71_41604,Gaunt2013Phys.Rev.Lett.110_200406},
we will show that the quantum coherence, originally absent in classical
thermodynamics, will in turn improve our understanding towards classical
thermodynamics in a particular example of entropy increasing.

The entropy of an isolated system is believed to monotonically increase
due to the irreversible process \cite{Huang1987EditionNewYork:JohnWiley&Sonsunknown_}.
However, a completely isolated system with no interaction with other
degrees of freedoms would follow a unitary evolution \cite{Breuer2002book}.
Such unitary evolution results in a constant entropy of the system,
namely $S\left(t\right)=S\left(0\right)$ where $S\left(t\right)=-k_{B}\mathrm{Tr}[\rho\left(t\right)\ln\rho\left(t\right)]$
and $\rho\left(t\right)=U\left(t\right)\rho\left(0\right)U^{\dagger}\left(t\right)$.
This observation simply implies that the ``isolated'' system in
classical thermodynamics is not completely isolated. The underlying
question is, to what extent, the system is isolated to retain the
statement of entropy increase in classical thermodynamics, while consistent
with other assumptions.

One of the mostly referred models for entropy increase is the free
expansion (FE) of ideal gas in a box, known as the Joule expansion
\cite{Callen1998unknown_,Goussard1993Am.J.Phys.61_845}. During FE,
the internal energy is kept constant, while entropy would increase
with $\Delta S_{c}=Nk_{B}\ln V_{A}/V_{B}$, where $V_{B}$ and $V_{A}$
are the volume of gas before and after expansion respectively \cite{Callen1998unknown_}.
The increase of the entropy is attributed to an irreversible process,
whose physical correspondence, to our best knowledge, remains vague.
In this letter, we show that such irreversible process could be a
pure dephasing process, by which the quantum coherence are essentially
removed after expansion. 

\begin{figure}
\includegraphics[width=1\columnwidth]{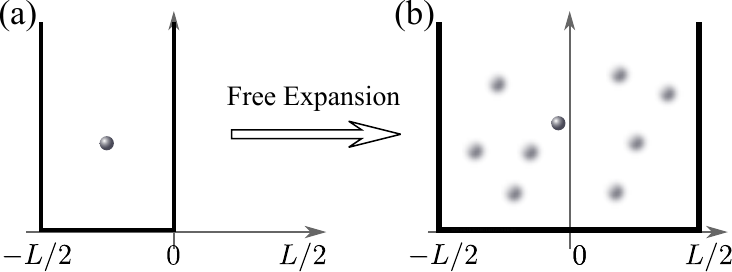}

\includegraphics[width=1\columnwidth]{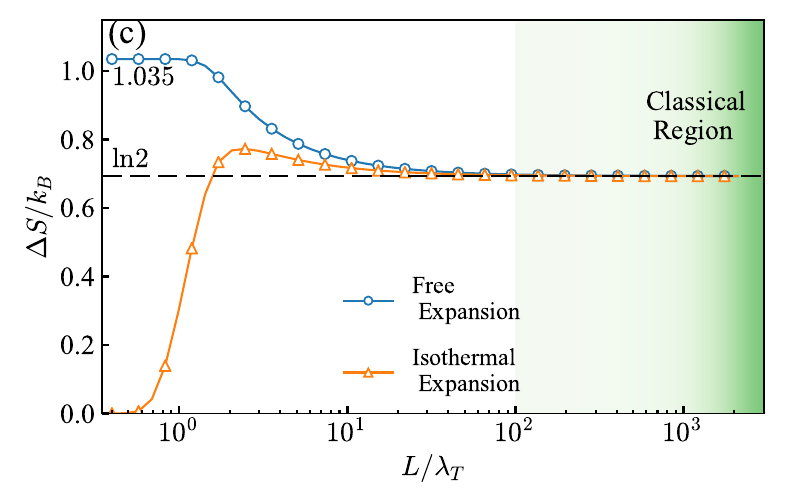}

\caption{(Color online). (a) General setup of the model of a single atom trapped
in a one dimensional square potential. The atom is initially trapped
on the left side with trap size $L/2$, and then expanded into the
whole trap with size $L$ after the free expansion. (b) Entropy change
($\Delta S$) during free (blue circle line) and isothermal (orange
triangle line) expansion as a function of ratio $\left(L/\lambda_{T}\right)$
between trap size and thermal wavelength of atom. The dashed gray
line shows the classical case of entropy change $S_{c}=k_{B}\ln2$.
The gray area shows the region where entropy change is almost equal
to the classical one. In this calculation, we increase the trap size
$L$, while with the temperature $T$ fixed. }

\label{fig1}
\end{figure}

To simplify the discussion, we consider a widely used model with a
single atom ( mass $M$) trapped in a 1D square potential \cite{Dong2011Phys.Rev.E83_61108,Cai2012Phys.Rev.E85_31114,Gong2016Phys.Rev.Lett.117_180603}
at temperature $T$, illustrated in Fig. \ref{fig1}(a). We include
a quantum dephasing effect, which conserves the energy in order to
match Joule's statement of no energy exchange during FE in classical
thermodynamics. With this dephasing process, the entropy change during
FE, is shown as a function of the ratio between the trap size $L$
and the thermal de Broglie wavelength $\lambda_{T}=h(2\pi Mk_{B}T)^{-1/2}$
with the orange curve in Fig. \ref{fig1}(b). For comparison, we also
plot the entropy change of an isothermal expansion (IsoE), where the
atom, in contact with a thermal bath, has a constant temperature.
At the classic region ($L/\lambda_{T}\gg1$), the entropy changes
for both processes match the well-known classical result $\Delta S_{c}=k_{B}\ln2$,
when the trap size doubles. Apart from the classic region, e.g. $L/\lambda_{T}<1$,
the entropy change after FE deviates from the classical one $\Delta S_{c}$.
At the limit $L/\lambda_{T}=0$, the entropy changes for two processes
are different: it approaches a constant $1.035k_{B}$ for FE, while
reaches zero for IsoE. In the IsoE process, the atom is frozen to
the ground state with both entropy and total energy constantly reduced
as the temperature decreases.

\textit{Theoretical framework} - Before expansion, the atom is trapped
on the left side by a half-size square potential $V_{b}\left(x\right)$,
\begin{equation}
V_{b}\left(x\right)=\begin{cases}
0 & -L/2<x<0\\
\infty & \mathrm{otherwise}
\end{cases},
\end{equation}
as shown in Fig. \ref{fig1}(a). For the single-atom Hamiltonian $H_{b}=p^{2}/2M+V_{b}\left(x\right)$,
the eigen-wavefunction is denoted as $\left|\phi_{n}^{L}\right\rangle $
with the corresponding energy $E_{n}^{L}=4n^{2}\alpha$, where $\alpha=\pi^{2}\hbar^{2}/[2ML^{2}]$.
In the region $-L/2<x<0$, the probability amplitude of the atom is
$\left\langle x\right.\left|\phi_{n}^{L}\right\rangle =\sqrt{4/L}\sin(2\pi nx/L)$.
We assume the atom is initially in a thermal equilibrium state with
inverse temperature $\beta=1/(k_{B}T)$, 
\[
\rho\left(t_{0}\right)=\sum_{n=1}^{\infty}\frac{\exp[-\beta E_{n}^{L}]}{Z}\left|\phi_{n}^{L}\right\rangle \left\langle \phi_{n}^{L}\right|,
\]
where $Z=\sum_{n=1}^{\infty}\exp[-\beta E_{n}^{L}]\equiv\sum_{n=1}^{\infty}\exp[-qn^{2}]$
is the partition function with $q=4\alpha\beta$. The dimensionless
parameter $q$ is rewritten in terms of the trap size $L$ and the
thermal wavelength $\lambda_{T}$ as $q=2\pi(\lambda_{T}/L)^{2}$.
The partition function is simplified with Theta-function as $Z=[\theta_{3}\left(0,e^{-q}\right)-1]/2$.
The internal energy of atom is $\left\langle H_{b}\right\rangle =-\partial\ln Z/\partial\beta$,
which approaches $k_{B}T/2$ at high temperature limit \cite{Dong2012Sci.Chin.Phys.Mech.Astr.55_1727},
matching the equipartition theorem in classical thermodynamics \cite{Huang1987EditionNewYork:JohnWiley&Sonsunknown_}.

After a sudden moving of the right wall, we have a new trap potential
$V_{a}\left(x\right)$ with twice the size $L$. The corresponding
eigen wavefunction for the new Hamiltonian $H_{a}=p^{2}/2m+V_{a}\left(x\right)$
is denoted as $\left|\psi_{n}\right\rangle $ with energy $E_{n}=n^{2}\alpha$.
The initial state of atom can be rewritten with this new basis $\{\left|\psi_{n}\right\rangle \}$,

\begin{equation}
\rho\left(t_{0}\right)=\sum_{m}D_{m}\left|\psi_{m}\right\rangle \left\langle \psi_{m}\right|+\sum_{m_{1}\neq m_{2}}F_{m_{1}m_{2}}\left|\psi_{m_{1}}\right\rangle \left\langle \psi_{m_{2}}\right|,
\end{equation}
where $D_{m}=\sum_{n=1}^{\infty}\left\langle \psi_{m}\right.\left|\phi_{n}^{L}\right\rangle \left\langle \phi_{n}^{L}\right.\left|\psi_{m}\right\rangle \exp[-qn^{2}]/Z$
is the diagonal term, and $F_{m_{1}m_{2}}=\sum_{n=1}^{\infty}\left\langle \psi_{m_{1}}\right.\left|\phi_{n}^{L}\right\rangle \left\langle \phi_{n}^{L}\right.\left|\psi_{m_{2}}\right\rangle \exp[-qn^{2}]/Z$
is the off-diagonal element, corresponding to the quantum coherence. 

\begin{figure}
\includegraphics[width=1\columnwidth]{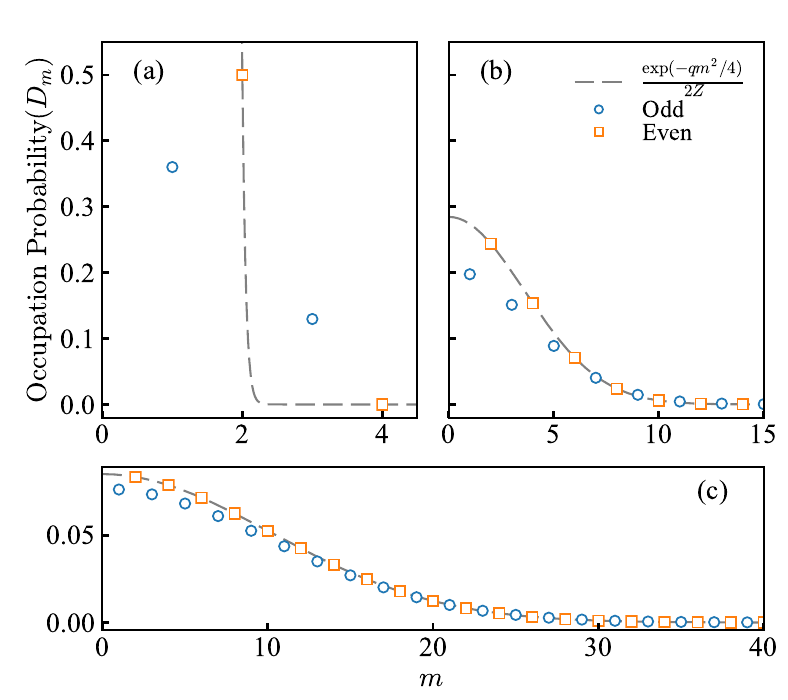}

\caption{Distributions of occupation numbers $D_{m}$ at different temperatures:
(a) $T=1$ (b) $T=100$ (c) $T=1000$. The corresponding thermal distributions
are plotted as gray dashed lines on each sub-figures.}

\label{fig:distribution}
\end{figure}

Clearly, the off-diagonal term is relevant in counting the entropy
at least for the initial state, namely, $\mathrm{Tr}[\rho_{0}\ln\rho_{0}]\neq\sum D_{m}\ln D_{m}$.
If the atom is completely isolated with no interaction with any other
degrees of freedom, this non-vanishing coherence term would contribute
to entropy, resulting in no entropy change. With this observation,
we remark here that a dephasing mechanism would be necessary for demolishing
the coherence in order to recover the general statement of entropy
increase in thermodynamics. 

The immediate question is whether depashing solely is sufficient to
recover the classical result of entropy increase. The answer is yes.
Supposing only dephasing is involved to demolish all off-diagonal
elements, we obtain a completely mixed state $\rho_{\mathrm{f}}=\sum_{m}D_{m}\left|\psi_{m}\right\rangle \left\langle \psi_{m}\right|$.
In the following discussion, we will concentrate on the discussion
on the effect of quantum dephasing, while delay the corresponding
dynamics to next section.

Let's first calculate the diagonal terms $D_{m}$. The probabilities
for even and odd quantum numbers are
\begin{equation}
D_{m}=\begin{cases}
\frac{\exp[-qm^{2}/4]}{2Z}, & m=2k\\
\sum_{n=1}^{\infty}\frac{32n^{2}\exp[-qn^{2}]}{Z(m^{2}-4n^{2})^{2}\pi^{2}}, & m=2k-1
\end{cases}\label{eq:distribution}
\end{equation}
with $k=1,2,3...$. We remark that the summation in $D_{m}$ can be
analytically performed for even number, yet not for odd number. The
analytic simplicity for even number is a direct result of the double
size after expansion. The nodes of wavefunctions after expansion match
the boundary of the wall position before expansion. One interesting
result is the odd and even portions have equal total contributions,
namely, $\sum_{k=1}D_{2k}=\sum_{k=0}D_{2k+1}=1/2$. 

We show the probability $D_{m}$ in Eq. (\ref{eq:distribution}) for
both even and odd numbersin Fig. \ref{fig:distribution} at $T=1$
(Fig.\ref{fig:distribution}a), $T=100$ (Fig. \ref{fig:distribution}b),
and $T=1000$ (Fig. \ref{fig:distribution}c). Gray dashed lines mark
the probabilities at thermal equilibrium with initial temperatures,
except for a normalization factor $1/(2Z)$. The distribution of even
number states indeed follows that of thermal equilibrium at the initial
temperature, as in Fig. \ref{fig:distribution}, while the distribution
for the odd-number state deviates from the thermal distribution at
initial temperature. Therefore, the atom after FE is not on a thermal
equilibrium. Such deviation from the thermal equilibrium is reduced
at high temperature, illustrated in Fig. \ref{fig:distribution}(c).

The FE process conserves the total energy of gas atom. The effect
of conserving energy is directly reflected through an extreme case
at zero temperature. At zero temperature, the initial state before
expansion is the ground state $\left|\phi_{n=1}^{L}\right\rangle $
with energy $E_{\mathrm{ini}}=4\alpha$, which is four times of the
energy of the ground state $\left|\psi_{n=1}\right\rangle $ of the
gas after FE. Taking the limit, one directly gets $D_{2k}=1/2\delta_{k,1}$
$\left(k=1,2,3...\right)$, which implies atom has half probability
to go the even-number state with the same energy $E_{2}=E_{1}^{L}$.
The other half of initial internal energy is redistribution into odd
number states with the distribution $D_{2k-1}=32/[((2k-1)^{2}-4)^{2}\pi^{2}]$.

With the distribution above, the gas entropy after expansion is ready
to be calculated via von Neumann-Shannon entropy definition
\begin{equation}
S_{\mathrm{f}}=-k_{B}\mathrm{Tr}[\rho_{\mathrm{f}}\ln\rho_{\mathrm{f}}],
\end{equation}
while the entropy before expansion is $S_{\mathrm{i}}=-k_{B}\mathrm{Tr}\left[\rho_{\mathrm{i}}\ln\rho_{\mathrm{i}}\right]$.
The entropy change after FE $\Delta S_{\mathrm{FE}}=S_{\mathrm{f}}-S_{\mathrm{i}}$
is plotted vs the ratio $L/\lambda_{T}$, as blue line with circles
in Fig. \ref{fig1}. In the calculation, we have used natural unit,
$k_{\mathrm{B}}=1$ and $\hbar=1$, and set the mass $M=1$, the temperature
$T=1$. In the simulation, the ratio $L/\lambda_{T}$ is changed via
varying the trap size $L$. The curve shows an asymptotic approaching
to the classical case with entropy change $\Delta S_{\mathrm{c}}=k_{B}\ln2$
at the classical region $L/\lambda_{T}\gg1$. This asymptotic behavior
confirms that the well-know classical result of entropy increase is
retained with solely considering dephasing. In this region, the entropy
change during IsoE is close to that of FE, as shown in Fig. \ref{fig1}(c).
However, the processes are different: (1) FE involves the dephasing
mechanism as the irreversible process, while IsoE is reversible. (2)
The atom's internal energy during FE is kept constant, while decreases
for IsoE process.

Notable difference at the region $(L/\lambda_{T}\sim1)$ is shown
in Fig. \ref{fig1}. The entropy change during FE is larger than the
classical result $k_{B}\ln2$. The interesting case is at zero temperature,
where the entropy of initial state is zero, namely, $S_{i}=0$. A
direct calculation of the entropy after FE gives $S_{\mathrm{f}}=1.035k_{B}$,
as illustrated in Fig. \ref{fig1}. The constant $1.035$ can be calculated
via the probabilities at zero temperature. During IsoE, the atom is
frozen to the ground state both initially and finally at zero temperature.
Therefore, the entropy change is simple zero at zero temperature.

\begin{figure}
\includegraphics[width=1\columnwidth]{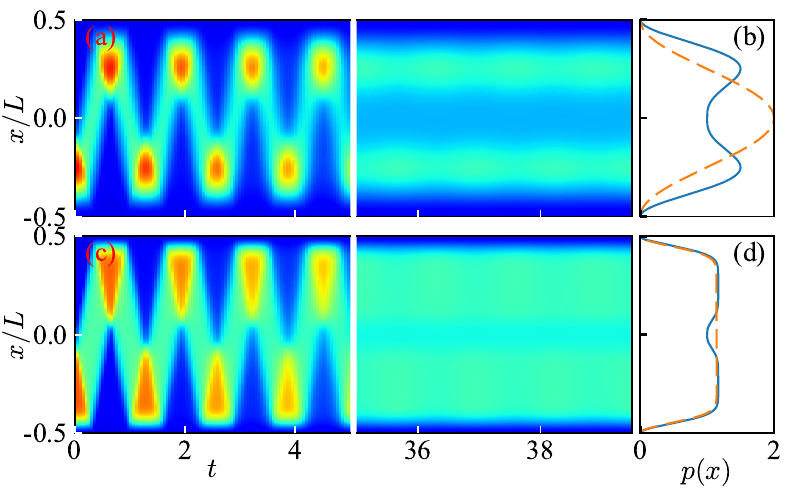}

\caption{Contour plot of density profile $p\left(x,t\right)$ at short time
$[0-5]$ and long time $[35-40]$ for both low $(T=1)$ and high ($T=100$)
temperature. The dephasing is set as the same, namely, $\gamma_{mn}=\gamma$. }

\label{fig:dynamics}
\end{figure}

\textit{Dephasing and Dynamics} \textendash{} By including dephasing,
we have shown the recovery of the classical result of entropy change,
and more importantly the deviation apart from classical region. We
now turn to the physical origin of the dephasing process, which can
be a result of the fluctuation of the wall of the trap. Supposing
the wall has small fluctuations, $L\rightarrow L+\delta L$, the eigen-energy
of the single particle is changed to $E_{n}\left(L+\delta L\right)\simeq E_{n}\left(L\right)+2E_{n}\left(L\right)\frac{\delta L}{L}.$
This coupling to the wall is similar to the coupling in cavity optomechanics,
where the cavity wall has a small displacement. Following the similar
procedure, we obtain an effective interaction between the trapped
atom and the trap wall as 
\begin{equation}
H_{int}=\sum_{n=1}^{\infty}g_{n}\left|\psi_{n}\right\rangle \left\langle \psi_{n}\right|\left(b+b^{\dagger}\right),
\end{equation}
where $b\left(b^{\dagger}\right)$ is the annihilation (creation)
operator for fluctuation of the trap wall, and $g_{n}=2\alpha n^{2}$.
Multiple modes of the wall with interaction $H_{int}=\sum_{n}\sum_{\xi}g_{n\xi}\left|\psi_{n}\right\rangle \left\langle \psi_{n}\right|\left(b_{\xi}+b_{\xi}^{\dagger}\right)$,
result in dephasing to gas molecule in the square trap, namely, 
\begin{align}
\frac{\partial}{\partial t}\rho\left(t\right) & =i\left[H,\rho\left(t\right)\right]+\mathcal{L}[\rho\left(t\right)],
\end{align}
where $\mathcal{L}[\rho\left(t\right)]=\sum_{mn}\gamma_{mn}\left(L_{m}^{\dagger}L_{n}\rho+\rho L_{m}^{\dagger}L_{n}-2L_{m}^{\dagger}\rho L_{n}\right)$
is the Liouville operator with $L_{n}=\left|\psi_{n}\right\rangle \left\langle \psi_{n}\right|$. 

In cold atom experiments, the density profile of gas $p\left(x,t\right)=\left\langle x\right|\rho\left(t\right)\left|x\right\rangle $
in the square potential is a measurable quantity. We show the dynamical
evolution of the density profile for both short time and long time
in Fig. \ref{fig:dynamics}(a) at low temperature $T=1$. In the simulation,
we set the same dephasing rate $\gamma_{nm}=\gamma$. The initial
localized atom on the left $\left(-1/2<x/L<0\right)$, flows to the
right and then bounces between the two walls, see Fig. \ref{fig:dynamics}(a)
at short time scale $t\in[0,5]$. Meanwhile, the atom spreads into
the whole square trap, illustrated by long-time behavior $t\in[35,40]$
in Fig. \ref{fig:dynamics}(a), with lowering revival peak on each
side as times of bouncing increases. The profile at the steady state
is illustrated in Fig. \ref{fig:dynamics}(b) with the blue solid
line, along with the profile of gas at equilibrium (the orange dashed
line). The profile after FE shows a distinct feature of a dip, illustrated
by the blue curve at middle of the square trap ($x/L=0$), instead
of a peak (orange line) for that at thermal equilibrium. The appearance
of the dip is caused by the non-equilibrium distribution shown in
Fig. \ref{fig:distribution}(a). The evolution of density profile
at higher temperature $\left(T=100\right)$ is shown in Fig. \ref{fig:dynamics}(c),
with the steady distribution in Fig. \ref{fig:dynamics}(d). At high
temperature, the discrepancy of steady-state profile from equilibrium
one is significantly reduced, as illustrated in Fig. \ref{fig:dynamics}(d).
The reduction of discrepancy from equilibrium one is the direct result
of distribution above.

\begin{figure}
\includegraphics[width=1\columnwidth]{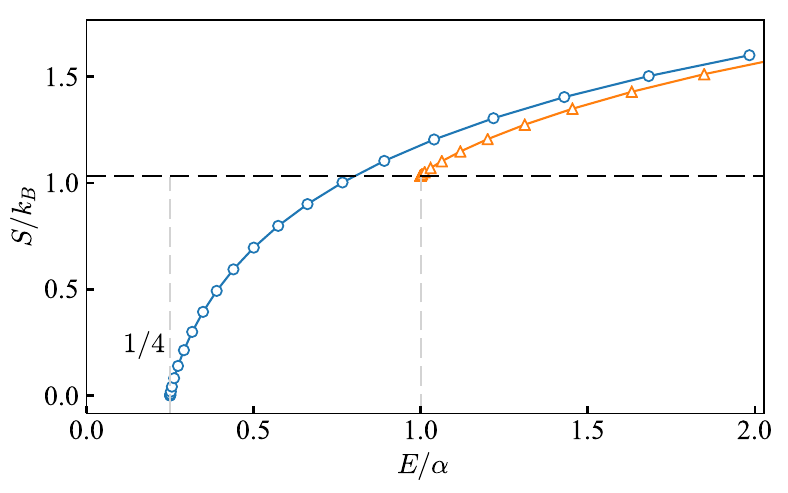}

\caption{Energy-entropy $(S(T)\sim E(T))$ relations for gas after free (orange
triangle curve) and isothermal (blue circle curve) expansion. The
horizontal gray dashed line marks the position of the constant $\mathcal{C}=1.035$.
In the calculation, we fix the size of the square trap $(L)$, and
increase the temperature.}

\label{fig:entropy-energy}
\end{figure}

One important relation in trapped atoms is the dependence of entropy
$S(T)$ on the internal energy $E(T)$ , namely $S(T)\sim E(T)$ curve
\cite{Luo2007Phys.Rev.Lett.98_80402}. We have shown that the atom
is at non-equilibrium state after free expansion. Such non-equilibrium
system has an abnormal $S-E$ curve, illustrated as the orange line
in Fig. \ref{fig:entropy-energy}. In the simulation, we increase
the temperature $\left(T\right)$ with the size $\left(L\right)$
of trap fixed. Due to the thermal isolation, the internal energy for
gas after free expansion is four times its ground state energy ($\pi^{2}\hbar^{2}/(2mL^{2})$)
at absolute zero temperature, while it is the ground state energy
for isothermal expansion. The resulting non-equilibrium distribution
after free expansion, leads to the non-zero entropy $S_{\mathrm{f}}\left(T\rightarrow0\right)=1.035k_{B}$,
illustrated in Fig. \ref{fig:entropy-energy} at zero temperature.
The corresponding of entropy-energy for gas at equilibrium is illustrated
as blue curve in Fig. \ref{fig:entropy-energy}. Different from the
curve for atom after FE, this curve shows both zero entropy and exact
ground state energy at zero temperature.

It's worthy to check the current experimental accessibility \cite{Gaunt2013Phys.Rev.Lett.110_200406}
for later verification of our theoretical observations. In Ref. \cite{Gaunt2013Phys.Rev.Lett.110_200406},
a square trap is experimentally realized with size $70\mathrm{\mu m}\times50\mathrm{\mu m}\times35\mathrm{\mu m}$,
and a Rubidium 87 atom with $120\mathrm{nk}$ has a thermal wavelength
$\lambda_{T}\sim1\mu m$. The ratio between trap size and thermal
wavelength is roughly $L/\lambda_{T}\sim40$, which is marked as an
arrow on Fig. \ref{fig1}. With these parameters, the difference from
the classical result is merely seen. To reach the region with prominent
difference, one needs to reduce the ratio by 10 times, by either reducing
the trap size to roughly several $\mathrm{\mu m}$ or the temperature
around $1\mathrm{nk}$. Such requirements is achievable, noticing
the coldest atom ensemble has been created in a sub picokelvin region
\cite{Kovachy2015Phys.Rev.Lett.114_143004}.

In conclusion, we have shown how the dephasing process in a quantum
fashion can serve as the irreversible process, and further deepen
our understanding of entropy increase in classical thermodynamics.
Furthermore, the discrepancies from classical thermodynamics are also
illustrated for the single atom in the square trap, far apart from
the classical region. We prove that such discrepancies are mainly
caused by the non-equilibrium state of the atom after FE, and can
be experimentally tested via normal measurements of the density profile
or energy-entropy relation.

H.D. would like to thank Marlan Scully for stimulating discussions
and valuable comments, and thank Lida Zhang,  Xiwen Zhang and Dazhi Xu for helpful suggestions. 
We gratefully acknowledge
support of the National Science Foundation Grant EEC-0540832 (MIRTHE
ERC), the Office of Naval Research (Award No. N00014-16-1-3054), and
Robert A. Welch Foundation (Grant No. A-1261).

\bibliographystyle{apsrev4-1}
\bibliography{freeexpansion}

\end{document}